\documentclass[pra,print,superscriptaddress,twocolumn]{revtex4}
\usepackage[dvips]{graphics,color}
\usepackage{mathrsfs}
\usepackage{amsfonts}
\usepackage{amsmath}
\usepackage{amssymb}
\usepackage{graphicx}
\usepackage{multirow}
\usepackage{color}
\usepackage[colorlinks,citecolor=blue]{hyperref}
\usepackage{times}

\makeatletter

\newcommand{\Rmnum}[1]{\expandafter\@slowromancap\romannumeral #1@}
\makeatother

\begin{document}

\title{Interplay between topology and localization on superconducting circuits}
\author{Xin Guan}
\affiliation{Department of Materials and Chemical Engineering, Taiyuan University, Taiyuan, DC 030032, China}
\email{guanxin810712@163.com}
\author{Gang Chen}
\affiliation{State Key Laboratory of Quantum Optics and Quantum Optics Devices, Institute of Laser Spectroscopy, Shanxi University, Taiyuan, DC 030006, China}
\affiliation{Collaborative Innovation Center of Extreme Optics, Shanxi University, Taiyuan, DC 030006, China}
\affiliation{School of Physics, Zhengzhou University, Zhengzhou, DC 450001, China}

\begin{abstract}
Topological insulator lie at the forefront of condensed matter physics. However strong disorder can destroy the topological states and make all states become localized. In this paper, we investigate the competition between topology and localization in the one-dimensional Su-Schrieffer-Heeger (SSH) model with controllable off-diagonal quasi-periodic modulations on superconducting circuits. By utilizing external ac magnetic fluxes, each transmon can be driven and all coupling strengths can be tuned independently. Based on this model we construct phase diagrams that illustrate the extended topologically nontrivial, critical localization, and coexisting topological and critical localization phases. The dynamics of the qubits' excitations are also discussed in this paper, revealing distinct quantum state transfers resulting from the interplay between topology and localization. Furthermore, we propose a method for detecting different quantum phases using current experimental setups. 
\end{abstract}

\maketitle

\section{Introduction}

Recent experimental breakthrough of superconducting circuits has opened up a
new avenue for processing quantum computing\cite{BI,GM,liu2018QCOE} and quantum information\cite{WG,yang2018QIOL,blais2020quantum}, as well
as implementing quantum simulation\cite{georgescu2014quantum,zhang2023superconducting,ying2023experimental}. Due to the superconducting circuits can
be fabricated into different lattice structures, a wide range of condensed
matter phenomena can be simulated, such as chiral physics and spin clusters\cite{Spincluster},
topological semimetal\cite{2018topologicalsemimential,toposemimental}, quantum walks\cite{walk,GM}, quantum phase transitions\cite{021spinphase,2022Phasetransition}, and many-body quench dynamics\cite{zhang2023superconducting}. A recent experiment
has successfully engineered a chain of qubits on superconducting circuits with tunable couplings through external drivings\cite{Xue2018perfect}.  Utilizing this experimental setup, a meticulous study has successfully observed the topological magnon insulator and detected its winding number\cite{Mei2019magnon}. These experiments provide a novel approach for investigating exotic
topological properties.

On the other hand, topological insulators(TIs)\cite{Insulator2010,qi2011topological,khanikaev2013TIs,rechtsman2013TIs,maciejko2015TIs,asboth2016TIs,rachel2018TIs,schindler2018TIs,tokura2019TIs,rudner2020TIs,biesenthal2022TIs,lei2022second} have received considerable attention
in the past decades. As is well known, in addition to possessing a bulk gap, topological insulators (TIs) exhibit symmetry-protected gapless states localized at their boundaries. The topological properties of TIs are robust against weak disorders such as fluctuations, imperfections in device fabrications, and environmental noises. Consequently, TIs have found numerous applications in quantum computation and information processing. However, when subjected to strong enough disorder, the topological state can be destroyed and replaced by a localization state. Additionally, moderate levels of disorder can induce a topological Anderson insulator\cite{li2009TAI,groth2009TAI,guo2010TAI,stutzer2018TAI,meier2018TAI,yang2020TAIOE,zhang2021TAI}. It is intriguing to investigate the interplay between topology and localization. Quasi-periodic disorder and dimer lattice structure have attracted much attention as means to induce localization and topological states, respectively. The famous Aubry-Andr\'{e}(AA)\cite{sokoloff1985AAH,thouless1988AAH,roushan2017AAHSCQ,an2021AAH,ye2022AAH,yang2022AAH} and Su-Schrieffer-Heeger(SSH)\cite{su1979SSH,marchand2010SSH,meier2016SSH,xing2021SSH,zhang2021SSHOE,kiczynski2022SSH,youssefi2022SSHSQC,wang2023SSH,wei2023SSHOE} models are abstracted, both of which have been realized in superconducting circuits\cite{roushan2017AAHSCQ,youssefi2022SSHSQC}. Therefore to design a qubits chain with staggered
coupling strength and controllable off-diagonal quasi-periodic modulations
is appealing.

In this paper, we propose a practical approach to realize the one-dimensional SSH model on superconducting circuits with controllable off-diagonal quasi-periodic modulations. Specifically each transmon is driven by an external ac magnetic flux
via the flux-bias line. All parameters can be independently adjusted by manipulating the amplitudes and frequencies of the drivings. Through this generalized SSH model, we have discovered extended topologically nontrivial, critical localization, and co-existing topological and critical localization phases transitions. We utilize the winding number and inverse participation ratio (IPR) to characterize the topological and localization properties, respectively. Additionally, we discuss the dynamics of qubit excitations in this paper. We observe different quantum state transfers induced by competition between topology and localization. Finally, we address possible experimental observations of these quantum phases. Herein, we introduce the mean chiral displacement (MCD) as a means to probe the winding number. Our results provide a novel avenue for exploring exotic quantum phases and dynamic behaviors arising from the interplay between dimerization and quasi-periodic disorder strengths on superconducting circuits.

\section{Model and Hamiltonian}
\begin{figure}[h!]
	\centering
	\includegraphics[height=4cm]{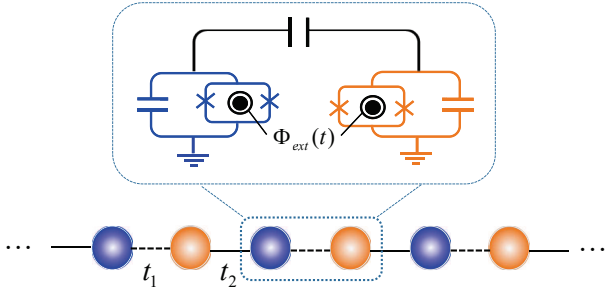}\newline
	\caption{ Schematic diagram of transmon qubits chain with tunable coupling strengths. All transmon qubits are
		coupled with their nearest-neighbor qubits by capacitors. (a) Illustration of
		the detail circuits of two coupled qubits.  (b) Schematic diagram
		of the generalized SSH model with $t_1$ and $t_2$ as the coupling strengths. $\Phi_{ext}(t)$ is the external ac magnetic flux.   }
	\label{QC}
\end{figure}
The method for implementing a one-dimensional SSH model with quasi-periodicity on superconducting circuits is as follow. We consider a chain of transmon qubits that are capacitively coupled to their next-nearest neighbor qubits, as illustrated in Fig. \ref{QC}(a). The corresponding Lagrangian is written as 
\begin{equation}
	\mathcal{L}=\underset{j}{\sum }\left[ \frac{C_{j}}{2}\dot{\phi}%
	_{j}^{2}+E_{j}^{J0}\cos \left( \frac{\phi _{j}}{\phi _{0}}\right) \right] +%
	\underset{\left\langle i,j\right\rangle }{\sum }\left[ \frac{C_{ij}}{2}%
	\left( \dot{\phi}_{i}-\dot{\phi}_{j}\right) ^{2}\right]   \label{L}
\end{equation}%
Here, $C_{j}$ and $E_{j}^{J0}$ represent the effective capacitance and the
Josephson energy of the $j$th transmon qubit. $C_{ij}$ denotes the capacitor used to
couple the $i$th and $j$th transmon qubits with its nearest-neighbor qubits. 
$\phi _{j}$ refers to node flux expressed in units of reduced flux quantum $%
\phi _{0}=1/\left( 2e\right) $. The symbol $\left\langle i,j\right\rangle $
indicates that the summation runs over all pairs of the nearest-neighbor
sites. With the canonical coordinate $\hat{\phi}_{j}$, the canonical
momentum $\hat{q}_{j}=\partial L/\partial \dot{\hat{\phi}}_{j}$, i.e.,%
\begin{equation}
	\hat{q}_{j}=C_{j}\dot{\hat{\phi}}_{j}+\underset{j\left\langle i\right\rangle 
	}{\sum }C_{j}\left( \dot{\hat{\phi}}_{j}-\dot{\hat{\phi}}_{j\left\langle
		i\right\rangle }\right) 
\end{equation}%
where $j\left\langle i\right\rangle $ is the summation of the
nearest-neighbor sites around the $j$th site. Considering the experimental
condition that $C_{j}\gg \{C_{\left( j-1\right) j},C_{j\left( j+1\right) }\}$%
, we have $\dot{\hat{\phi}}_{j}\approx \hat{q}_{j}/C_{j}$. By applying the
Legendre transformation, we obtain the corresponding Hamiltonian%
\begin{equation}
	\hat{H}=\hat{H}_{\text{0}}+\hat{H}_{\text{IJ}},  \label{TH}
\end{equation}%
where%
\begin{equation}
	\hat{H}_{\text{0}}=\underset{j}{\sum }\left( \frac{\hat{q}_{j}^{2}}{2\tilde{C%
		}_{j}}+\frac{\hat{\phi}_{j}^{2}}{2L_{j}^{J}}-\frac{E_{j}^{J0}}{24\hat{\phi}%
		_{0}^{4}}\hat{\phi}_{j}^{4}\right) ,
\end{equation}%
\begin{equation}
	\hat{H}_{\text{IJ}}=\underset{\left\langle i,j\right\rangle }{\sum }\frac{%
		C_{ij}}{C_{i}C_{j}}\hat{q}_{i}\hat{q}_{j},
\end{equation}%
with $1/\tilde{C}_{j}=(C_{j}-C_{\left( j-1\right) j}-C_{j\left( j+1\right)
})/C_{j}^{2}$ and $1/L_{j}^{J}=E_{j}^{J0}/\phi _{0}^{2}$.

By introducing bosonic annihilation and creation operators 
\begin{eqnarray}
	\hat{a}_{j} &=&\sqrt{\frac{\tilde{C}_{j}\omega _{j}^{0}}{2}}\hat{\phi}_{j}+i%
	\sqrt{\frac{1}{2\tilde{C}_{j}\omega _{j}^{0}}}\hat{q}_{j}, \\
	\hat{a}_{j}^{\dag } &=&\sqrt{\frac{\tilde{C}_{j}\omega _{j}^{0}}{2}}\hat{\phi%
	}_{j}-i\sqrt{\frac{1}{2\tilde{C}_{j}\omega _{j}^{0}}}\hat{q}_{j},
\end{eqnarray}%
where $[\hat{a}_{j},\hat{a}_{j^{\prime }}^{\dag }]=\delta _{j,j^{\prime }}$
and $\omega _{j}^{0}=\sqrt{8E_{j}^{C}E_{j}^{J0}}$ with $E_{j}^{C}=e^{2}/(2%
\tilde{C}_{j})$, and after applying a rotating-wave approximation, the
Hamiltonian (\ref{TH}) is quantized as 
\begin{eqnarray}
	\hat{H}_{\text{0}} &=&\sum_{j}\omega _{j}^{0}\hat{a}_{j}^{\dag }\hat{a}_{j}+%
	\frac{1}{2}V_{j}\hat{n}_{j}(\hat{n}_{j}-1),  \label{H0} \\
	\hat{H}_{\text{IJ}} &=&\sum_{j}g_{j}^{0}\hat{a}_{(j-1)}^{\dag }\hat{a}_{j}+%
	\text{H.c.},  \label{HIJ}
\end{eqnarray}%
where H.c.~is the Hermitian conjugate, $\hat{n}_{j}=\hat{a}_{j}^{\dag }\hat{a%
}_{j}$ is the particle number operator, and 
\begin{eqnarray}
	V_{j} &=&-E_{j}^{C},  \label{V} \\
	g_{j}^{0} &=&\frac{C_{\left( j-1\right) j}\sqrt{\omega _{\left( j-1\right)
			}^{0}\omega _{j}^{0}}}{2\sqrt{C_{\left( j-1\right) }C_{j}}}.  \label{g1}
\end{eqnarray}%
In Eqs.~(\ref{H0}) and (\ref{HIJ}), $V_{j}$\ denotes the anharmonicity of
each qubit, $g_{j}^{0}$ is the coupling strength between the $j$th and ($j-1)
$th qubits. Since the requirement of $E_{j}^{C}\ll E_{j}^{J0}$ for transmon qubits\cite{koch2007Transmon}, we choose $E_{j}^{J0}/E_{j}^{C}=50$, $%
\{C_{\left( j-1\right) j},C_{j\left( j+1\right) }\}\approx 0.5$ fF, and $%
C_{j}\approx 100$ fF. Based on these parameters, we obtain $V_{j}/\left( 2\pi
\right) \approx 200$ MHz and $g_{j}^{0}/\left( 2\pi \right) \approx 10$ MHz\cite{ye2019two24}.
This results in a condition that $V_{j}\gg g_{j}^{0}$. In such case, the qubit
exhibits strong anharmonicity, allowing for only one boson to be excited per transmon qubit. This results in $V_{j}\hat{n}_{j}(\hat{n}_{j}-1)/2\equiv 0$,
and $\hat{a}_{j}^{\dag }\left\vert 0\right\rangle _{j}=\left\vert
1\right\rangle _{j}$, $\hat{a}_{j}^{\dag }\left\vert 1\right\rangle _{j}=0$, 
$\hat{a}_{j}\left\vert 0\right\rangle _{j}=0$, and $\hat{a}_{j}\left\vert
1\right\rangle _{j}=\left\vert 0\right\rangle _{j}$. As a result, the
Hamiltonian (\ref{TH}) becomes%
\begin{equation}
	\hat{H}=\sum_{j}\omega _{j}^{0}\hat{a}_{j}^{\dag }\hat{a}_{j}+\sum_{j}\left(
	g_{j}\hat{a}_{(j-1)}^{\dag }\hat{a}_{j}+\text{H.c.}\right).   \label{HM}
\end{equation}%
The commutation relation can be reformulated as follows,
\begin{equation}
	\lbrack \hat{a}_{j},\hat{a}_{j^{\prime }}^{\dag }]=(1-2\hat{n}_{j})\delta
	_{j,j^{\prime }}.  \label{HCO}
\end{equation}

To achieve tunable coupling strength, an ac microwave driving is applied to
each transmon qubit, which can be implemented experimentally via the flux-bias line 
line\cite{Xue2018perfect}. As illustrated in Fig. \ref{QC}, we incorporate a Josephson junction into a superconducting quantum interference device (SQUID).
Since the two Josephson junctions in SQUID possess equal energy, the effective energy can be expressed as  $E_{j}^{J}=2E_{j}^{J0}$. 
As a result of the external magnetic flux $\Phi _{j}(t)$, a phase difference $\Phi _{j}(t)/\phi
_{0}$ arises between the two Josephson junctions, leading to a modification in the potential energy of each qubit,%
\begin{eqnarray}
	&&E_{j}^{J0}\left[ \cos \left( \frac{\phi _{j}}{\phi _{0}}+\frac{\Phi _{j}(t)%
	}{2\phi _{0}}\right) +\cos \left( \frac{\phi _{j}}{\phi _{0}}-\frac{\Phi
		_{j}(t)}{2\phi _{0}}\right) \right]   \notag \\
	&=&E_{j}^{J}(t)\cos \left( \frac{\phi _{j}}{\phi _{0}}\right) ,
\end{eqnarray}%
where $E_{j}^{J}(t)=E_{j}^{J}\cos [\Phi _{j}(t)/2\phi _{0}]$. The frequency
of the $j$th qubit becomes $\omega _{j}(t)=\omega _{j}^{0}\sqrt{\cos [\Phi
	_{j}(t)/2\phi _{0}]}.$ Here we take the external magnetic flux as,%
\begin{equation}
	\Phi _{j}(t)=\alpha _{j}\cos (u_{j}^{0}t+\phi _{j}^{0}),
\end{equation}%
where $\alpha _{j}\ll \omega _{j}^{0}$, $u_{j}^{0}$, and $\phi _{j}^{0}$ are
amplitude, frequency, and phase of the external magnetic flux respectively.
Next we conduct Taylor expansion of the frequency $\omega _{j}(t)$ at $\Phi
_{j}(t)=0$,%
\begin{equation}
	\omega _{j}(t)\approx \omega _{j}^{0}+\varepsilon _{j}\sin (u_{j}t+\phi
	_{j}),
\end{equation}%
where $\varepsilon _{j}=\omega _{j}^{0}\alpha _{j}^{2}/4\phi _{0}^{2}$, $%
u_{j}=2u_{j}^{0}$, and $\phi _{j}=2\phi _{j}^{0}+3\pi /2.$ In experiments\cite{Xue2018perfect}, $%
u_{j}$ is of the order $\sim 100$ MHZ, which is far larger than $g_{j}^{0}$. Then an
unitary operator $\hat{U}=\hat{U}_{A}\times \hat{U}_{B}$ is introduced to
transform the Hamiltonian (\ref{HM}) with the frequency $\omega _{j}(t),$
where,%
\begin{eqnarray}
	\hat{U}_{A} &=&\exp \left[ -i\underset{j=1}{\overset{N}{\sum }}\omega
	_{j}^{0}\hat{n}_{j}t\right]   \notag \\
	\hat{U}_{B} &=&\exp \left[ i\underset{j=1}{\overset{N}{\sum }}f_{j}\cos
	(u_{j}t+\phi _{j})\right] ,
\end{eqnarray}%
with $f_{j}=\varepsilon _{j}/u_{j}$ and $N$ is the number of the qubits.
The transformed Hamiltonian can be obtained,%
\begin{align}
	\hat{H}_{t} =&\hat{U}^{\dag }\hat{H}\hat{U}+i\frac{d\hat{U}^{\dag }}{dt}%
	\hat{U}  \notag \\
	=&\overset{N-1}{\underset{j=1}{\sum }}g_{j}^{0}\left\{ e^{-i\Delta_{j}t}\exp [-if_{j}\cos (u_{j}t+\phi _{j})]\right.\notag\\
	&\left.\exp [if_{j+1}\cos (u_{j+1}t+\phi
	_{j+1})]\hat{a}_{j}^{\dag }\hat{a}_{j+1}+\text{H.c.}\right\},  \label{HH}
\end{align}%
where $\Delta _{j}=\omega _{j+1}^{0}-\omega _{j}^{0}.$
We choose $\Delta _{j}=u_{j+1}(-u_{j+1})$ for odd (even) $j$.
Next we utilize the Jacobi-Anger identity, $\exp [if_{j}\cos (u_{j}t+\phi
_{j})]=\sum_{-\infty }^{\infty }i^{m}J_{m}(f_{j})\exp [im(u_{j}t+\phi _{j})],
$ and apply the
rotating-wave approximation to the Hamiltonian in Eq. (\ref{HH}). Here $J_{m}(f_{j})$ represents the $m$th Bessel function of the first kind.  By neglecting
the oscillating terms and setting $\phi _{j}=-\pi (0)$ for odd (even) $j$,
we can derive,%
\begin{equation}
	\hat{H}_{eff}=\overset{N-1}{\underset{j=1}{\sum }}g_{j}\left( \hat{a}%
	_{j}^{\dag }\hat{a}_{j+1}+\text{H.c.}\right) ,  \label{HSSH}
\end{equation}%
where 
\begin{equation}
	g_{j}=g_{j}^{0}J_{0}(f_{j})J_{1}(f_{j+1})
\end{equation}%
As is evident, the coupling strength $g_{j}$ can be tuned independently. For
obtaining the SSH model with controllable off-diagonal quasi-periodic modulations, we take $g_{j}=g[1+\lambda
+\delta \cos (2\pi \beta j)]$ for odd $j$ and $g_{j}=g[1-\lambda +\delta
\cos (2\pi \beta j)]$ for even $j$. For constituting the quasi-periodic
disorder, we take $\beta =(\sqrt{5}-1)/2$. $\lambda $ and $\delta $ indicate
the dimerization and quasi-periodic disorder strengths respectively. In this
paper we set $g=1$ as the unit of the energy. In the following discussion,
we denote the creation (annihilation) operator of the qubit with odd and
even $j$ as $\hat{a}_{i}^{\dag }(\hat{a}_{i})$ and $\hat{b}_{i}^{\dag }(\hat{%
	b}_{i})$ respectively, where $i$ is the unit cell index in the case of $\delta=0$. Here we consider
the single excitation throughout our work.

\section{Phase diagram and quantum state transfer}

\subsection{Phase diagram}
\begin{figure}[h!]
	\centering 
	\includegraphics[height=7cm]{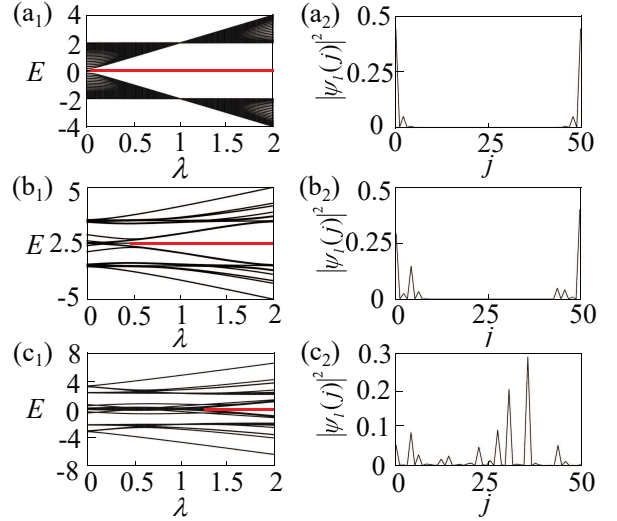}\newline
	\caption{ Energy bands and wave functions of the $N/2$th and $(N/2+1)$th$[l=N/2,(N/2+1)]$ eigenstates for the Hamiltonian in Eq. (\ref{HSSH}) with open boundary condition.  ($a_{1}$)-($c_{1}$) Energy bands as functions of $\lambda$ with ($a_{1}$) $\delta=0$, ($b_{1}$) $\delta=1$, and ($c_{1}$) $\delta=2$. ($a_{2}$)-($c_{2}$) Wave functions with ($a_{2}$) $\delta=0$ and $\lambda=0.5$, ($b_{2}$) $\delta=1$ and $\lambda=0.5$ and ($c_{2}$) $\delta=2$ and $\lambda=0.5$.     }
	\label{EB}
\end{figure}

In this section we focus on the phase diagram resulting from the interplay between quasi-periodic disorder and dimerization strengths. We employ 50 transmon qubits in this section.

When disorder is absent in the generalized SSH model (\ref{HSSH}), it is
reduced to the SSH model. Its energy bands are shown in Fig. \ref{EB}(a),
indicating the presence of degenerate zero-energy modes that persist within the 
middle of the energy gap, with a value of $\lambda >0$. The corresponding wave functions
exhibit exponential localization at the boundaries of the system, as shown in Fig. \ref{EB}%
(e). Subsequently, we activate the disorder and present the variation of these energy bands in Figs. \ref%
{EB}(b) and (c), demonstrating that as $\delta $ increases, the
energy gap near zero-energy vanishes within the small $\lambda $ region.
Figures \ref{EB} (e)-(g) demonstrate the transition of edge states into
bulk of the system as $\delta $ increases. These results demonstrate that strong disorder can only destroy topological states in the case of small $\lambda $. This implies that the computation between dimerization strength $\lambda $ and quasi-periodic disorder strength $\delta $ can induce a transition from topologically trivial to nontrivial phases. In the following, we will explain these results analytically.

With $\delta =0,$ we can ansatz the edge states as $\left\vert \Psi
\right\rangle _{E=0}=\sum_{i}(\rho _{1}^{i}w_{1}\hat{a}_{i}^{\dag }\pm \rho
_{2}^{i}w_{2}\hat{b}_{i}^{\dag })\left\vert G\right\rangle $, where $%
\left\vert G\right\rangle $ is the vacuum state. According to the Schr\"{o}%
dinger equation, $\hat{H}\left\vert \Psi \right\rangle _{E=0}=0$, ie,%
\begin{equation}
	(g_{1}\rho _{2}^{i}+g_{2}\rho _{2}^{i-1})w_{2}\hat{a}_{i}^{\dag }\left\vert
	G\right\rangle +(g_{2}\rho _{1}^{i+1}+g_{1}\rho _{1}^{i})w_{1}\hat{b}%
	_{i}^{\dag }\left\vert G\right\rangle =0,
\end{equation}%
the edge states can be obtained as%
\begin{equation}
	\left\vert \Psi \right\rangle _{E=0}=\underset{i}{{\sum }}\left[
	\left( -\frac{g_{1}}{g_{2}}\right) ^{i}w_{1}\hat{a}_{i}^{\dag }\pm \left( -%
	\frac{g_{2}}{g_{1}}\right) ^{i}w_{2}\hat{b}_{i}^{\dag }\right] \left\vert
	G\right\rangle .\label{state}
\end{equation}%
These results demonstrate that the state is
localized in proximity to the leftmost ($i=1$) $a-$type site and rightmost ($i=N/2$) $b-$%
type site.

Next, we introduce disorder and observe the changes in $\rho _{1}$ and $\rho _{2}$,%
\begin{eqnarray}
	\rho _{1} &=&-\frac{1-\lambda +\delta \cos \left[ 2\pi \beta \left(
		2i-1\right) \right] }{1+\lambda +\delta \cos \left[ 2\pi \beta \left(
		2i\right) \right] },  \label{rho1} \\
	\rho _{2} &=&-\frac{1+\lambda +\delta \cos \left[ 2\pi \beta \left(
		2i\right) \right] }{1-\lambda +\delta \cos \left[ 2\pi \beta \left(
		2i-1\right) \right] },  \label{rho2}
\end{eqnarray}%
which depend on the qubits' indexes. Equations (\ref{rho1}) and (\ref{rho2}) indicates the presence of non-zero $\delta$ prevents $\rho_{1}$ and $\rho_{2}$ from always being larger or smaller than 1. This implies that the behavior of topological edge states is influenced by  $\delta$. Then for characterizing distinct topological phases in the case of $\delta\neq0$, we compute the topological invariant.
Since the Hamiltonian (\ref{HSSH}) possesses chiral, time-reversal and
particle-hole symmetries, it belongs to class BDI of the Altland--Zirnbauer
classification. The topological properties of the system can be characterized by the winding number, which can be easily calculated in momentum space for translationally invariant systems. However, in the presence of non-zero disorder strength, translational symmetry is broken and thus the computation of winding number must be performed in real space using a local quantity\cite{mondragon2014winding,meier2018winding},%
\begin{equation}
	\nu =\frac{1}{2}\text{Tr}^{\prime }(\hat{P}\hat{\Gamma}\hat{X}\hat{P}),\label{WW}
\end{equation}%
where Tr$^{\prime }$ indicate the trace over the single unit cell in the
middle of the qubits chian. $\hat{P}=\hat{P}_{+}-\hat{P}_{-}$ with $\hat{P}%
_{+}(\hat{P}_{-})$ is the projection operator of the eigenstates with
positive (negative) eigenenergies. $\hat{\Gamma}=\sum_{i=1}^{N}\left( \hat{%
	\Gamma}_{a_{i}}^{e}-\hat{\Gamma}_{b_{i}}^{e}\right) $ where $\hat{\Gamma}%
_{di}^{e}=$ $\left\vert e\right\rangle _{di}\left\langle e\right\vert $ with 
$\left\vert e\right\rangle _{di}$ ($d=a,b$) indicates the state with
exciting $d-$type qubit at the $i$th unit cell$.$ $\hat{X}$ is the unit-cell
position operator. We numerically calculate the winding number in Fig. \ref%
{PD}(a). In region \uppercase\expandafter{\romannumeral3} ($\delta >1+\lambda $), $\nu =0$, indicating that
the system is topologically trivial. In regions \uppercase\expandafter{\romannumeral1} and \uppercase\expandafter{\romannumeral2} ($\delta <1+\lambda 
$), $\nu =1$, indicating that the system is topologically nontrivial. This finding accurately demonstrates that the coexistence of strong disorder and weak dimerization strength can result in the annihilation of topological states.
\begin{figure}[h!]
	\centering
	\includegraphics[height=4cm]{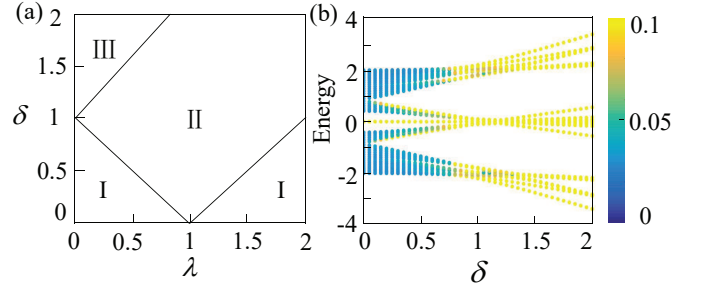}\newline
	\caption{ (a) Phase diagram in the $\lambda$-$\delta$ plane. \uppercase\expandafter{\romannumeral1} is extended topologically nontrivial phase with winding number $\nu=1$. \uppercase\expandafter{\romannumeral2} is the coexisting topological critical localization phase with winding number $\nu=1$. \uppercase\expandafter{\romannumeral3} is the critical localization phase with winding number $\nu=0$. (b) Energy bands denoted by the values of IPR$_n$ as a function of $\delta$ with $\lambda=0.5$.  }
	\label{PD}
\end{figure}

On the other hand, localization always emerges in a quasi-periodic system.
We can capture the localization property of the $n$th eigenstate $\left\vert
\Psi _{n}\right\rangle =\sum_{j}\psi _{n}(j)\left\vert e\right\rangle _{j}$ (%
$\psi _{n}(j)$ is the wave function) in our system by the inverse
participation ratio(IPR), which is defined as IPR$_{n}=\sum_{j}\left\vert
\psi _{n}(j)\right\vert ^{4}$. For an extended (a fully localized) state, IPR%
$_{n}\approx (2N)^{-1}$ (IPR$_{n}\approx $1). For a critically localized
state, IPR$_{n}\approx (2N)^{-\theta },$where $\theta \in (0,1)$ depends on
the multifractal nature of the wave function. We present the phase diagram in
Fig. \ref{PD}(a) and find our system exhibits localization transition
boundaries, where the values of IPR$_{n}$ shift from zero in region
\uppercase\expandafter{\romannumeral1} ($\delta <1-\lambda $ $\cup $ $\delta <\lambda -1$) to finite values
between 0 and 1 in regions \uppercase\expandafter{\romannumeral2} and \uppercase\expandafter{\romannumeral3} ($\delta >1-\lambda \cap \delta >\lambda
-1$). This result suggests that the localization properties are also influenced by the strength of dimerization.

In conclusion, the competition between $\delta$ and $\lambda$ induces four different quantum phases which are extended topologically nontrivial [region \uppercase\expandafter{\romannumeral1} in Fig. \ref{PD}(a)], coexisting topological and critical localization [region \uppercase\expandafter{\romannumeral2}in Fig. \ref{PD}(a)], and critical localization [region \uppercase\expandafter{\romannumeral3} in Fig. \ref{PD}(a)] phases. In addition, the energy bands with open boundary condition
marked by the values of IPR$_{n}$ are shown in Fig. \ref{PD}(b), which
indicates there are no mobility edges except few edge states. 

\subsection{Quantum state transfer}
\begin{figure}[h!]
	\centering
	\includegraphics[height=8cm]{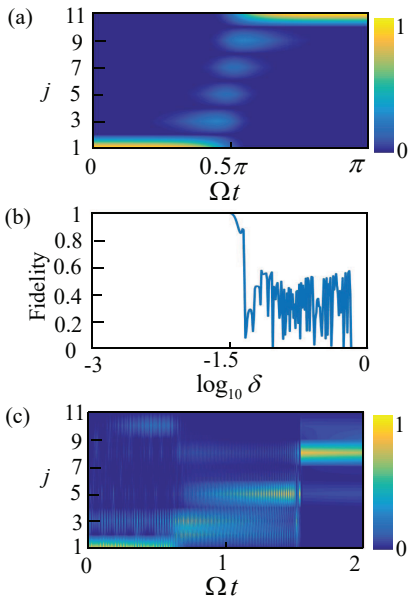}\newline
	\caption{ (a)Time evolution of qubits' excitations with $\delta=0$ and $L=11$. (b)  The fidelity of the state transfer between $\left| L\right\rangle$ and $\left| R\right\rangle$. (c) Time evolution of qubits' excited states with $\lambda=0.5$ and $L=10$ }
	\label{QST}
\end{figure}
In this section, we demonstrate the transfer of quantum states both with and without quasi-periodic disorder. 

Firstly, considering $\delta=0$, the edge states are presented in Eq.(\ref{state}). When the number of qubits is even, the edge states will localize at both boundaries of the system. Therefore, regardless of how $\lambda$ changes, the quantum state localized at leftmost boundary cannot transfer to the rightmost. While in the case of odd qubits, both boundaries are composed of $a$-type qubits which result in the localization of edge states at the left- and rightmost boundaries with $\lambda>0$ and $\lambda<0$, respectively. This implies the quantum state transfer can occur when $\lambda$ changes from negative to positive values. To achieve this change, we set $\lambda=\cos(\gamma)$ with $\gamma\in[0,\pi]$. In order to verify this quantum state transfer, the periodic parameter $\gamma$ should be a function of time $t$. Here we choose $\gamma=\Omega t$ with $\Omega$ as the rate of change and the initial state is chosen as the leftmost edge state, which can be written as $\left| L\right\rangle =(egg...g)$. We illustrate the temporal evolution of the qubits' excitations in Fig.\ref{QST}(a). This result indicates that the quantum state is transferred from the leftmost boundary to the rightmost boundary of the system through bulk with $\gamma$ varying from 0 to 2$\pi$. Subsequently, we introduce quasi-periodic disorder $\delta$ and present the fidelity of this quantum state transfer $F=\left\langle R\right| \psi _{f}\rangle$ in Fig.\ref{QST}(b). $\left| R\right\rangle=(gg..ge)$ is the rightmost edge state. $\left| \psi _{f}\right\rangle$ is the final state. When $\delta$ is sufficiently small, the quantum state transfer persists. While as $\delta$ increases beyond a certain threshold, the transfer becomes unfeasible.  

We then observe the quantum state transfer by varying the quasi-periodic disorder strength $\delta$. Similarly to the aforementioned discussion, here we set $\delta=\Omega t$ and choose $\left| L\right\rangle$ as the initial state. The corresponding temporal evolutions of the qubits' excitations are illustrated in Fig. \ref{QST}(c). We find the qubits' excitations localize at the leftmost boundary when $\Omega t$ is smaller than approximately 0.5, which can be attributed to the system being in a topological extended phase as depicted in Fig. \ref{PD}(a). As $\Omega t$ varies from 0.5 to 1.5, these excitations depart from the leftmost boundary and diffuse throughout the bulk of the system. In this case, the system is in coexisting topological and critical localization phase [Fig. \ref{PD}(a)]. Therefore this diffusion arises due to competition between topology and localization. In the region where $\Omega t>1.5$, the qubits' excitations are localized at one or some bulk states of the system due to critical localization phase with topologically trivial properties[Fig. \ref{PD}(a)]. These results suggest that the quantum state can transfer from a boundary state to certain states within the bulk, which is induced by the interplay between topology and localization.

\section{Possible experimental observation}

In this section, we propose an experimental scheme for detecting the
aforementioned phases. Furthermore, it is worth noting that the quantum state transfer can be observed directly.
\begin{figure}[h!]
	\centering
	\includegraphics[width=8cm]{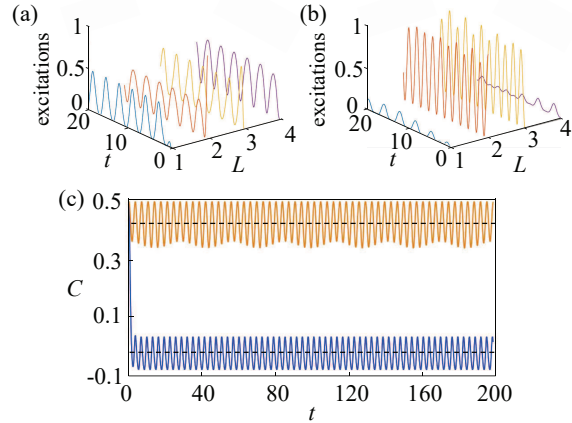}\newline
	\caption{ Time evolution of all qubits' excited states in two dimensional representation with (a) $\delta=2$ and (b) $\delta=0.2 $. (c) Time evolution of chiral displacement with $\delta=2 $ and $\delta=0.2 $ denoted by blue and orange curves, respectively. In all figures $\lambda=0.5$}
	\label{MDC}
\end{figure}
In superconducting circuits, the winding number (\ref{WW}) can be dynamically detected though the mean chiral
displacement (MCD) $\overline{C}$ \cite{meier2016MCD,cardano2017MCD,meier2018winding} with $\overline{C}=\nu /2$. The MCD is
defined as,%
\begin{equation}
	\overline{C}=\frac{1}{T}\int_{0}^{T}Cdt,
\end{equation}%
with $C=\left\langle \psi (t)\right\vert \hat{\Gamma}\hat{X}\left\vert \psi
(t)\right\rangle $. Here $\left\vert \psi (t)\right\rangle =e^{-i\hat{H}%
	t}\left\vert \psi (0)\right\rangle $ is the state of the system at time $t.$ 
$\left\vert \psi (0)\right\rangle $ is the initial state. Here we utilize
four qubits to detect the MCD. Firstly we prepare an initial state $%
\left\vert \psi (0)\right\rangle =\left\vert gegg\right\rangle ,$ which
excite one of the middle qubits to a excited state and leave the others in
ground states. The time evolution of this excitation can be measured
directly, which is shown in Fig. \ref{MDC}(a) and (b). We can see, in the
topological trivial phase, this excitation can propagate to the boundaries
of the qubit chain[Fig. \ref{MDC}(a)], whereas in the topologically nontrivial
phase, it is confined within the bulk[Fig. \ref{MDC}(b)]. Subsequently, the time
evolution of the CD can be derived, which as depicted in Fig. \ref{MDC}(c). The
blue and orange curves denote topological trivial and nontrival phases
respectively. The CD exhibits oscillations around -0.01 and 0.42 in
topological trivial and nontrivial phases, respectively, indicating distinct
winding numbers of -0.02 and 0.84. This finding is in close proximity to the ideal values.

To identify the critical localization, the detection of IPR$_{n}$ is necessary. The IPR$_{n}$ can be obtained from the eigenstates of
Hamiltonian (\ref{HSSH}). Since each qubit can be represented as a point
on the Bloch sphere, the $j$th qubit can be denoted by $\left\vert \psi
_{j}\right\rangle =\cos (\theta _{j})\left\vert 0\right\rangle _{j}+e^{i\chi
	_{j}}\sin (\theta _{j}/2)\left\vert 1\right\rangle _{j}$, where $\theta _{j}$
($\chi _{j}$) is the angle between the qubit $\left\vert \psi
_{j}\right\rangle $ and the $z(x)$ axis in the Bloch sphere. The eigenstates can be described as $\left\vert \psi \right\rangle
_{n}=\prod_{j}\left\vert \psi _{j}\right\rangle .$ To determine the specific
values of the angles $\theta _{j}^{n}$ and $\chi _{j}^{n}$ for the $n$th
eigenstate, it is necessary to numerically calculate the amplitude of the
single-qubit excitation $\lambda _{j}^{n}$ for each qubit in the $n$th eigenstate. Then angles $\theta _{j}^{n}\ $and $\chi _{j}^{n}$ can be
obtained as $2\arccos (\sqrt{1-\left\vert \lambda _{j}^{n}\right\vert ^{2}})$
and $-i\ln [\lambda _{j}^{n}/\sin (\theta _{j}/2)]$ respectively. The $n$th
eigenstate can be obtained after applying a rotation operator $\hat{U}%
_{j}^{n}=e^{i\theta _{j}^{n}\hat{\sigma}_{z}+i\chi _{j}^{n}\hat{\sigma}_{x}}$
for each qubit. This rotation operation can be realized by a microwave
pules. Finally the IPR$_{n}$ can be derived directly. 

\section{Conclusion}

In summary, we have proposed a feasible experimental method to realize the one dimensional SSH model with
controllable off-diagonal quasi-periodic modulations on superconducting
circuits. For realizing independently tunable coupling strengths, we apply an external ac magnetic flux
to drive each transmon qubit. Then the generalized SSH model with both dimerization and quasi-periodic can be constructed. Based on this
model, extended topologically nontrivial, critical localization, and
coexisting topological and critical localization phases transitions are
discovered. We use the winding number and  IPR to
characterize the topological and localization properties respectively. The dynamics of the qubits' excitations are also discussed in this paper, revealing distinct quantum state transfers resulting from the interplay between topology and localization. Furthermore, we propose a method for detecting different quantum phases using current experimental setups. Notably, we introduce the mean chiral displacement (MCD) as a means to probe the winding number. Our results provide a novel avenue for exploring exotic quantum phases and dynamic behaviors arising from the interplay between dimerization and quasi-periodic disorder strengths on superconducting circuits.
\bibliographystyle{unsrtnat}
\bibliography{REFF2}
\end{document}